# 面向GWAC系统的流星候选体识别算法*

徐 洋[1,2]，王 竞[1,2,3]，黄茂海[1,2]，魏建彦[1,2]

（1. 中国科学院国家天文台空间天文与技术重点实验室，北京 100101；2. 中国科学院大学，北京 100049；
3. 广西大学物理科学与技术学院广西相对论天体物理重点实验室，广西 南宁 530004）

**摘要**：地基广角相机阵的超大视场每天可观测数百个流星体，有效识别这些流星体可以为流星的科学研究提供重要的科学数据。针对地基广角相机阵这一类超大视场测光巡天系统在流星识别时遇到的不能有效区分流星与其他移动目标的问题，设计和实现了一个流星候选体识别算法，该算法主要包含流星轨迹识别和光变曲线形态分析两部分。识别算法对Mini-GWAC约两个月的图像进行处理，提取10.9万个移动目标轨迹，其中90%以上属于非流星目标。分析午夜时间段内流星目标高斯拟合曲线的$\alpha$参数，发现大部分单峰流星目标的光变曲线波峰随图像像素位置呈慢速变化趋势。综合流星的单帧特性、光变曲线的单峰结构特征和光变曲线的慢速变化特性进行过滤，最终得到4.1%的高精度流星候选体。经过人工检查确认，4.1%的流星候选体中有85%~87.3%的目标符合流星的形态及亮度特征。

地基广角相机阵(Ground Wide Angle Cameras，GWAC)是中法合作的多波段空间天文变源监测器(Space-based multiband astronomical Variable Objects Monitor，SVOM)[1]的地基相关观测设备，主要科学目标是探测伽玛射线暴、引力波事件等爆发时的电磁辐射对应体。除了探测伽玛射线暴等暂现源，地基广角相机阵约5 000平方度的超大视场同时能够探测到大量的移动目标，如流星、近地目标、小行星、彗星、飞机、空间碎片、卫星等。这些移动目标会影响伽玛暴等暂现源的探测效率，识别并过滤这些移动目标对地基广角相机阵具有重要意义。同时，国际上许多组织在进行各类移动目标的研究，如国际天文联盟的流星数据中心致力于流星数据的收集和流星特性的研究[2]，地基广角相机阵将贡献大量的流星数据。本文主要针对地基广角相机阵的流星目标设计识别算法。

国际上有许多流星研究项目：美国国家航空航天局资助建立的全天流星监测项目(Cameras for Allsky Meteor Surveillance project，CAMS)项目[3]包含3个站点共60个录像相机，其目标是系统研究流星雨的特性及来源；日本的SonotaCo①包含25个站点共100个录像相机，已经向国际天文联盟的流星数据中心贡献了大量的流星雨数据；欧洲视频流星观测网(European viDeo Meteor Observation Network，ED-MONd)流星数据库[4]汇聚了欧洲多个流星观测网的数据；克罗地亚流星网(Croatian Meteor Network，CMN)[5]基于廉价的监控相机和普通台式机构建面向科普教育的流星观测系统；加拿大Western大学的南安大略流星网(Southern Ontario Meteor Network，SOMN)[6]主要关注厘米级流星体的探测；俄罗斯的微型超大快速光学瞬变源研究望远镜(Mini-Mega TORTORA，MMT)[7]与地基广角相机阵系统有类似的科学目标，即探测短时标光学暂现源，MMT包含900平方度浅视场和100平方度的深视场，白光0.1 s曝光极限星等约11 V，1.5年累计观测到约9万颗流星[8]。

---

* 基金项目：中国科学院空间科学战略性先导科技专项(XDA15052600)；中国科学院战略性先导科技专项(B类)(XDB23040000)资助.
收稿日期：2019-02-18；修订日期：2019-03-06
作者简介：徐 洋，男，硕士. 研究方向：天文图像处理. Email：yxu@nao.cas.cn

① A meteor shower catalog based on video observations in 2007-2008

现有项目使用的流星识别软件如 MeteorScan，MetRec [9]，UFO Capture②，AIM-IT [10]，Gural [11] 和 Vida [12] 等，主要面向流星巡天项目进行设计，其图像的主要特点是: (1) 极限星等低，视场中背景恒星的数目少，识别流星时干扰源少; (2) 帧频快，拍摄的流星体轨迹出现在多帧图像中，每帧图像包含一段轨迹，通过多段轨迹可计算流星体的角速度; (3) 多站点观测，流星巡天项目通常包含少则两个多则几十个观测站点，联合两个及以上站点的观测数据，可以计算流星体的位置、高度和线速度。流星巡天项目中流星识别的过程为: (1) 提取多帧图像中的轨迹线段，关联多个轨迹线段为一个目标; (2) 通过多站点数据，计算目标的位置; (3) 通过多个轨迹线段的位置计算目标的速度; (4) 通过目标的高度和速度信息对流星候选体进行过滤。文 [13] 使用改进的帧差法对图像中拉长轨迹很短的空间运动目标进行检测。

地基广角相机阵图像的特点是: (1) 极限星等高，图像中会有大量的背景恒星及移动目标; (2) 帧频慢，拍摄的流星出现在单帧图像中，通常表现为一段完整的拉长轨迹，无法计算流星体的速度; (3) 目前地基广角相机阵仅有一个观测站点，无法计算流星体的位置和高度。由于地基广角相机阵的图像和图像中流星的轨迹形态与现有流星巡天项目有本质的区别，因此现有流星识别软件或算法不能直接应用于地基广角相机阵的流星探测。

国际上与地基广角相机阵有类似科学目标的大视场测光巡天项目有 MMT [7] 和 Pi of the Sky [14]，这两个项目也没有可借鉴的经验: (1) MMT 的曝光时间为 0.1 s，其流星识别算法与流星巡天项目的类似; (2) Pi of the Sky 没有发表流星识别的相关研究成果。

地基广角相机阵系统的超大视场每天不仅能观测数百个流星体，同时也能观测几千个其他类型的移动目标，如流星、飞机、卫星等。图 1 为 Mini-GWAC 图像中各种不同类别的移动目标，除(a) 是流星候选体之外，其他为不同类别的干扰源。与其他移动目标相比，流星轨迹的典型特征为: (1) 持续时间短，普遍单帧出现; (2) 轨迹在形态上中间宽两端窄，在亮度上中间亮两端暗(火流星例外)，如图 2。

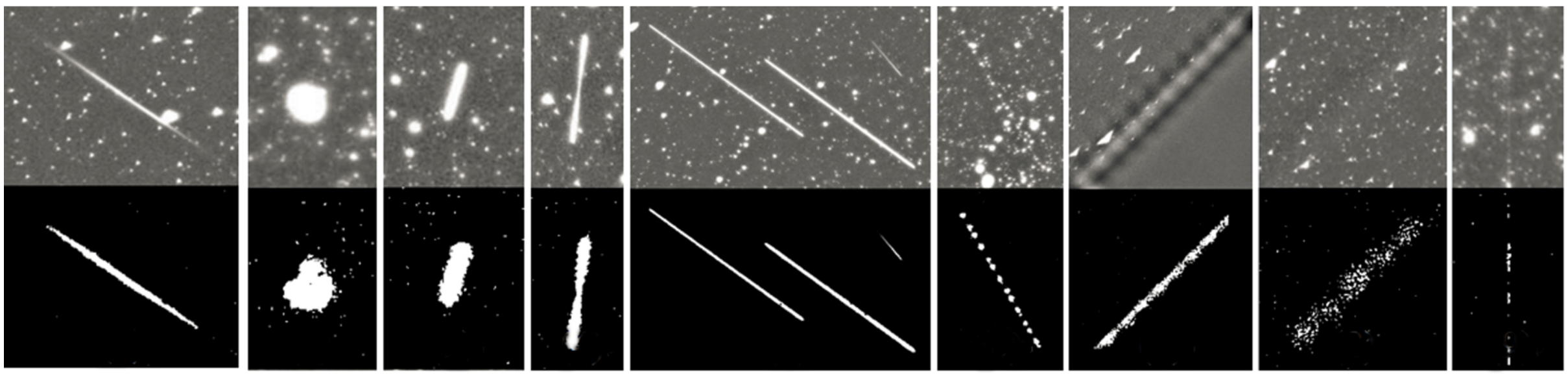

(a)　(b)　(c)　(d)　(e)　(f)　(g)　(h)　(i)

图 1　Mini-GWAC 图像中各种类别的目标，其中上面部分是原始图像，下面部分是相邻图像相减后的结果。(a) 流星候选体; (b~f) 不同类别的移动目标; (g) 望远镜视场边缘被遮挡后产生的假目标; (h) 流星的尾焰; (i) 热像列

Fig. 1　All kind of objects in Mini-GWAC, top part of image is original image, bottom part of image is residual image of adjacent image. (a) is meteor candidate, (b-f) are moving objects, (g) is fake object, (h) is wake flame of meteor, (i) is hot column

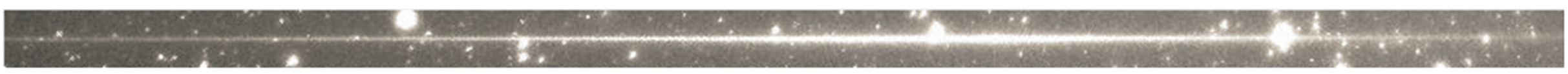

图 2　典型的流星示例，形态上中间宽两端窄，亮度上中间亮两端暗

Fig. 2　The center part of meteor is wider and brighter than both ends

地基广角相机阵的流星体识别过程面临的挑战是在无法依靠速度和高度信息准确区分流星体和其他移动目标的情况下，如何从大量的移动目标中找到流星体。针对地基广角相机阵的系统特性及其图像中流星的轨迹特征设计流星识别算法: (1) 针对流星的拉长轨迹和单帧特征设计流星轨迹识别算法; (2) 针对

---

② http://sonotaco.com/soft/UFO2/help/english/index.html

流星轨迹的亮度和形态特征，对流星候选体的形态特征进行分析，以进一步提高流星识别的准确率。

## 1 地基广角相机阵系统简介

### 1.1 地基广角相机阵观测系统

地基广角相机阵的建设包含两个阶段，分别是一期工程 Mini-GWAC 实验系统和二期工程地基广角相机阵，如图 3。

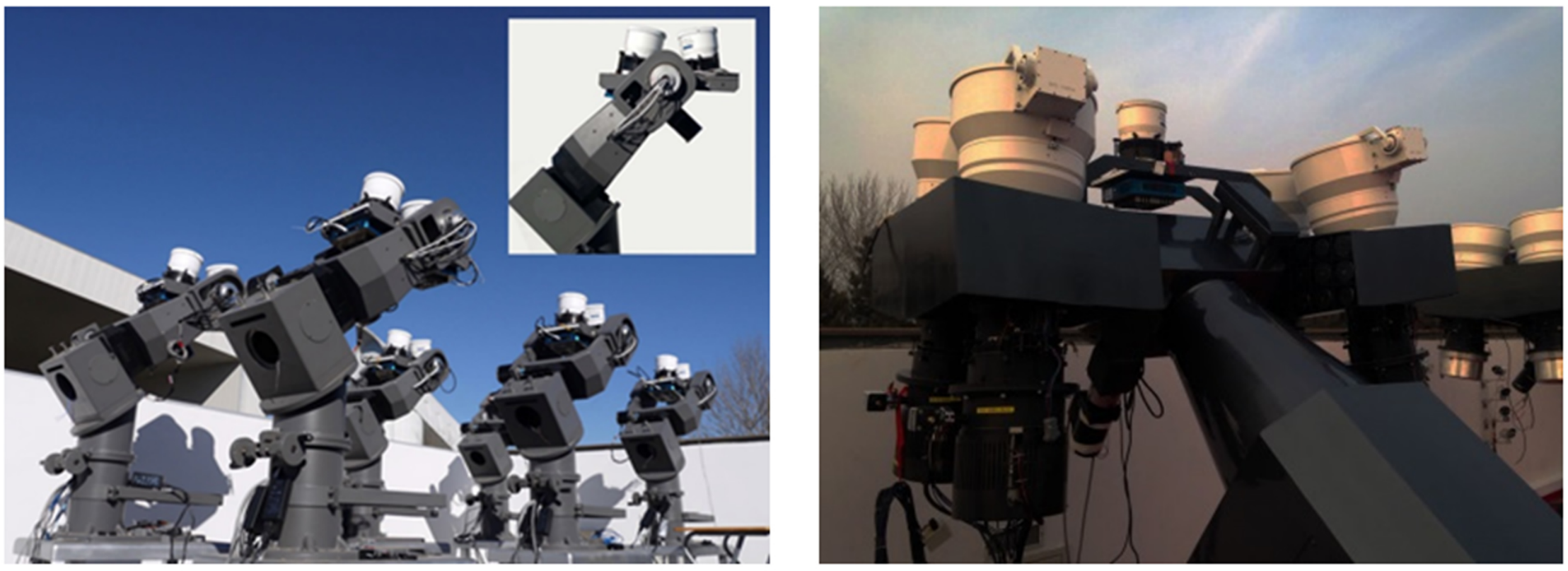

(a) (b)

图 3 地基广角相机阵系统。(a) Mini-GWAC; (b) GWAC

Fig. 3 Image on the left is Mini-GWAC, and on the right is GWAC

(1) Mini-GWAC 的组成：包含 12 台 6 cm 的宽视场望远镜，每台望远镜配备一台 3 k × 3 k 的 CCD 相机，综合视场约 5 000 平方度，恒星探测极限星等约 12.5 V，2014 年 10 月建成，并于 2015 年 10 月正式开始观测。

(2) 地基广角相机阵的组成：共包含 40 台有效口径为 18 cm 的宽视场望远镜，每台望远镜配备一台 4 k × 4 k 高性能 CCD 相机，综合视场约 5 000 平方度，恒星探测极限星等约 16.0 V。现在已完成部分望远镜的调试，并在进行试观测。

Mini-GWAC 已经积累了大量的观测数据，使用 Mini-GWAC 的数据进行流星识别算法的研究，并在后续应用于地基广角相机阵。

### 1.2 Mini-GWAC 系统移动目标轨迹特性分析

Mini-GWAC 图像中的移动目标主要包含 3 类：飞机、流星和卫星(空间碎片)，下面简要分析这 3 类目标的特性。

(1) 高度和速度特性

如表 1，飞机、流星和卫星 3 类移动目标的高度和速度区间有明显的差异，计算出高度或速度中的一项参数，即可区分移动目标的类别。地基广角相机阵目前为单站点观测，无法计算移动目标的高度或速度，因此无法利用高度或速度特性对移动目标进行区分。

表 1 飞机、流星和卫星的高度和线速度

Table 1 The height and speed of airplane, meteor and satellite

| | 最低高度 /km | 最高高度 /km | 最小线速度 /(km/s) | 最大线速度 /(km/s) |
|---|---|---|---|---|
| 飞机 | 6 | 12.6 | 0.08 | 0.3 |
| 流星 | 80 | 120 | 11 | 72 |
| 卫星 | 200 | 36 000 | 3 | 7.78 |

(2) 时间特性

MMT 与 Mini-GWAC 使用相同型号的镜头，1.5 年累计观测到约 9 万颗流星，这些流星持续时间为 0.1 s 到 2.5 s[8]。照此估算，流星在 Mini-GWAC 的图像(10 s 曝光，5 s 读出)中只能单帧出现。飞机、卫星等移动目标在飞出图像边缘前会持续存在，因此可以在多帧图像中连续出现。

(3) 轨迹长度特性

图 4 为对表 1 中 3 类典型移动目标在 Mini-GWAC 单帧 10 s 曝光图像中轨迹长度的估算。可见在不同速度和高度时，这 3 类移动目标在图像中的轨迹长度存在交集，因此无法直接通过轨迹长度对这 3 类移动目标进行区分。

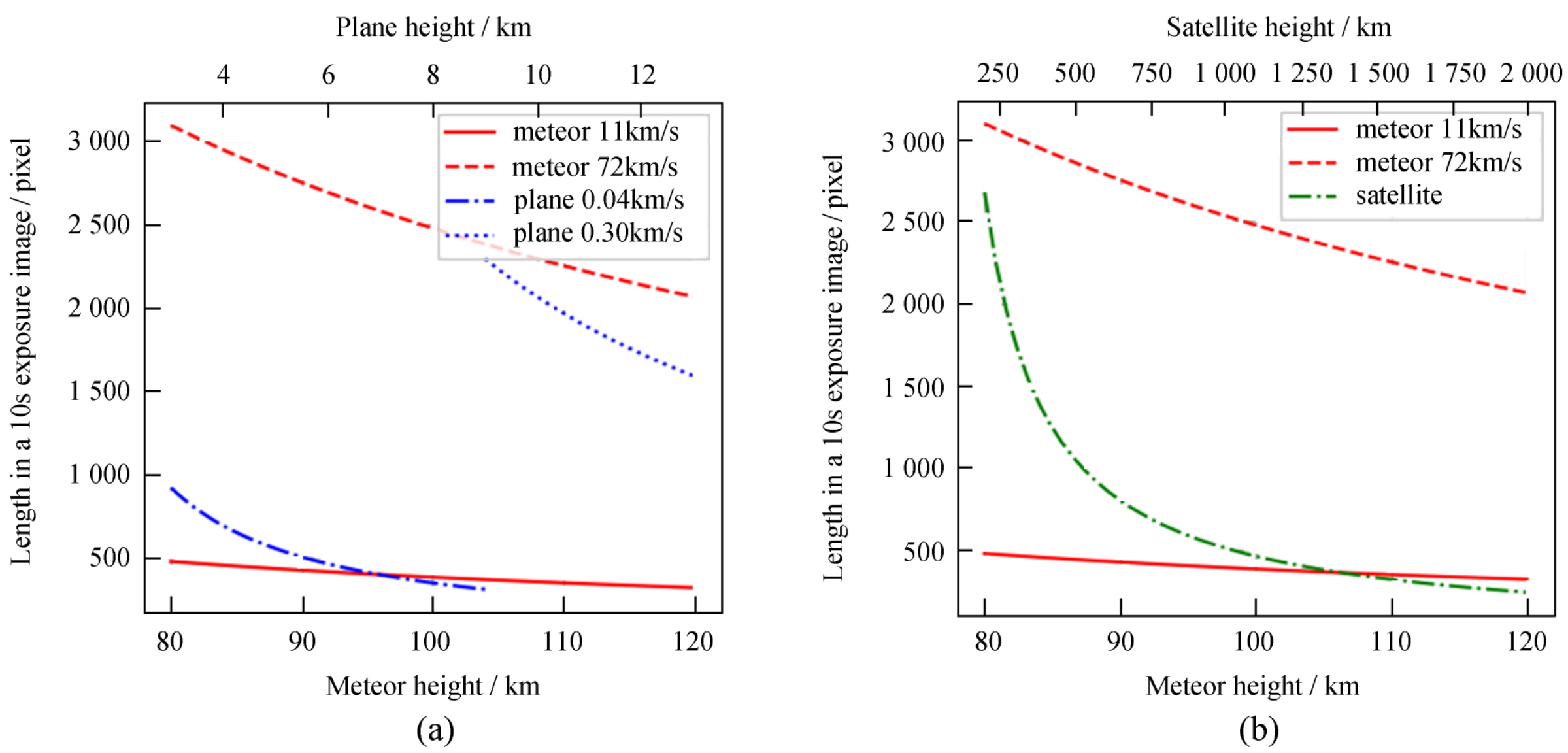

图 4　Mini-GWAC 的一帧 10 s 曝光图像中，流星、飞机和卫星的轨迹长度估算，其中流星的长度按 0.5 s 持续时间估算，飞机和卫星的长度按 10 s 持续时间估算。(a) 流星与飞机的轨迹长度对比；(b) 流星和卫星的轨迹长度对比

Fig. 4　The estimation of track length of airplane, meteor and satellite in a 10 seconds exposure image of Mini-GWAC, the left picture shows the track length comparison of the meteor and the plane, and the right picture shows the track length comparison of the meteor and the satellite

(4) 轨迹形态特性

文 [15] 对 CAMS 双站点观测的 891 颗流星进行了光变曲线和形状分析。(1) 光变曲线分析：光变曲线的形状有 67%的单峰目标，包括 14%的早峰类型，42%的对称类型，11%的晚峰类型；(2) 形状分析：单帧图像(110 帧/秒) 中流星为类彗星状。流星在 Mini-GWAC 图像中单帧出现，其轨迹也包含光变曲线和形状信息：(1) 光变曲线：从 Mini-GWAC 的图像中能够得到流星随位置变化的光变曲线；(2) 形状特征：Mini-GWAC 图像中流星的轨迹中间粗两端逐渐变细，其宽度仅为几像素到十几像素，采样不足，容易受噪声的影响，因此本文不对流星的形状进行分析。

## 2　流星候选体识别算法

### 2.1　流星轨迹识别

算法对时空连续的多张 FITS 图像 $I(k)$ 进行处理，整体流程如图 5，下面着重对几个关键的步骤进行介绍。

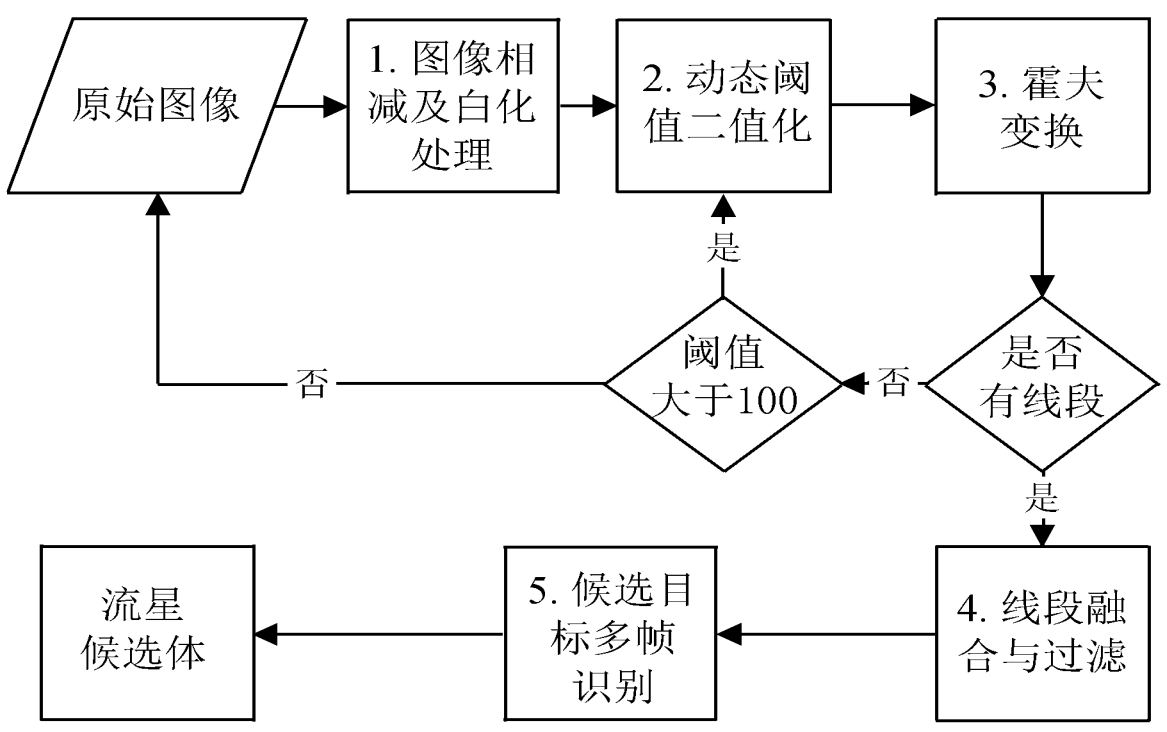

图 5　流星轨迹识别算法的流程图

Fig. 5　The flow chart of meteor track recognition method

(1) 图像预处理

连续的两帧图像拍摄时间相隔 15 s，图像之间视宁度等环境因素的变化非常微弱，可直接对连续两幅图像 $I(k)$ 和 $I(k-1)$ 进行相减操作，相减后残差图像 $S(k)$ 可表示为

$$S(k) = I(k) - I(k-1) \quad (k \text{ 为图像顺序编号，} k \in \mathbf{N} \text{ 且 } k \geqslant 2), \tag{1}$$

对相邻的两幅图像相减，避免了天文数据处理流程中对暗场和本底的操作，同时也不用对图像进行去背景操作。原始图像相邻像素点之间是有关联的，且相邻图像之间有细微的局部噪声起伏，为了降低环境因素对关键信息的影响，可对相减后的残差图像 $S(k)$ 进行白化处理。$S(k)$ 的协方差矩阵为 $R(k)$，白化处理后的图像 $W(k)$ 可表示为

$$W(k) = R(k)^{-1/2}S(k) \quad (k \in \mathbf{N} \text{ 且 } k \geqslant 2) . \tag{2}$$

相减后图像 $S(k)$ 主要包含 3 部分内容：背景、残差信号、真实信号。图像二值化过程包括：

1) 对 $W(k)$ 使用 $3\sigma$ 原则去掉背景信息，计算均值 $E$ 和标准差 $D$；

2) 初始化阈值设为 $T = E + 2D$，使用阈值 $T$ 对 $S(k)$ 进行二值化操作，得到二值化图像 $B(k)$：

$$B(k) = \begin{cases} 255, & W(k) \geqslant T \\ 0, & W(k) < T \end{cases} \quad (k \in \mathbf{N} \text{ 且 } k \geqslant 2) , \tag{3}$$

3) 如果霍夫变换操作没有得到线段集，且 $T > 100$，则开始动态调整阈值 $T$：

$$T = E + 2D - 5k \quad (k \in \mathbf{N} \text{ 且 } k \geqslant 1) . \tag{4}$$

其中，$k$ 为循环次数。每次循环将 $T$ 的值减 5，然后对图像进行二值化和线段提取操作。如果 $T < 100$ 或者(5)式找到线段集，则停止循环。

(2) 移动目标轨迹识别

选用概率霍夫变换(Progressive Probabilistic Hough Transform，PPHT)[14] 对二值化图像进行线段探测。相比霍夫变换，概率霍夫变换计算速度快，且能直接算出线段的两个端点坐标。概率霍夫变换后，得到候选线段集 $L(k)$ 可表示为

$$L(k) = PPHT[B(k)] \quad (k \in \mathbf{N} \text{ 且 } k \geqslant 2) . \tag{5}$$

流星在 Mini-GWAC 图像中的轨迹宽度约几像素到十几像素，因此霍夫变换后，一个流星轨迹有可能被识别为多条线段。为去除冗余线段，对候选线段集 $L(k)$ 进行融合，将相邻的多条线段合并为一条线段。综合采用线段中心距离、线段斜率和图像中心到线段的距离 3 个参数判断线段的相似性，对这 3 个参数小于指定数值的线段合并为一条线段 $L_0$，$L_0$ 代表一个移动目标候选体 $O_0$，对线段集 $L(k)$ 合并后得到移动目标候选集 $O(k)$。

(3) 多帧过滤

Mini-GWAC 图像中不仅包含流星，还包含其他的移动目标。流星与其他移动目标的区别是流星仅在单帧图像中出现，而其他移动目标可能在多帧图像中出现。因此，对合并后的移动目标候选集 $O(k)$ 进行前后帧关联，可以过滤一部分非流星目标。

### 2.2 流星光变曲线形态分析

移动目标光变曲线从形态上可分为无峰目标、单峰目标和多峰目标 3 类，如图 6。文[15]分析的流星样本中有 67%的流星是单峰目标，且其主要针对单峰目标进行光变曲线的形态分析。因此尝试对单峰目标的形态进行分析，以进一步过滤非流星目标。由于流星在 Mini-GWAC 图像中主要为单帧出现，无法得到流星随时间变化的光变曲线，但是可以通过流星在图像上的拉长轨迹得到流星随图像像素位置变化的光变曲线，后文简称光变曲线。下面简要介绍光变曲线的提取及单峰目标形态分析的关键步骤。

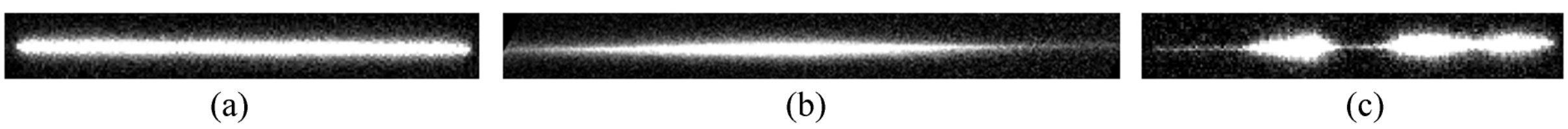

图 6 3 类移动目标。(a) 无峰目标；(b) 单峰目标；(c) 多峰目标

Fig. 6 Three kinds of moving objects. (a) is no-peak object; (b) is single-peak object; (c) is multi-peak object

(1) 光变曲线的提取

线段 $L_0$ 与原始图像 $x$ 轴的夹角为 $\theta_0$，对 $S(k)$ 顺时针旋转 $\theta_0$，使目标长轴在图像中处于水平位

置，对目标进行矩形区域裁图得到如图 6 的窗口图 $Sub_0$。对 $Sub_0$ 沿图像 $y$ 轴方向对 $x$ 轴方向的流量强度进行求和，得到目标亮度随 $x$ 的变化曲线:

$$Flux_x = \sum_y Sub_0(x, y) . \tag{6}$$

其中，$x$，$y$ 是图像沿 $x$ 轴和 $y$ 轴方向的坐标。

(2) 峰值计算及多峰目标过滤

峰值计算过程参考文 [17] 的多颗相邻恒星目标的解析过程。对光变曲线 $Flux_x$ 取与 $x$ 轴平行的直线 $y = Y_0$，$Y_0$ 从光变曲线 $Flux_0$ 的最大值开始，依次以 $Step_y$ 递减，直到最小值。每次递减后计算直线与光变曲线的交叉部分，交叉部分的个数即为峰值个数 $N_{peak}$。如果峰值个数 $N_{peak}$ 等于 0，则代表该目标为无峰目标; 如果峰值个数 $N_{peak}$ 等于 1，则代表该目标为单峰目标; 如果峰值个数 $N_{peak}$ 大于 1，则代表该目标为多峰目标。

(3) 光变曲线形态分析

采用高斯函数拟合的方式对光变曲线进行形状的量化，多个高斯函数叠加能更准确地描述目标的光变曲线，这里至多使用两个高斯函数对光变曲线进行拟合。光变曲线拟合函数如下:

$$Flux(k) = \sum_{i=0}^{N} A_i \mathrm{e}^{-\left(\frac{x - x_{0i}}{\alpha_i}\right)^2},\ N = [1, 2] , \tag{7}$$

其中，$A_i$，$x_{0i}$ 和 $\alpha_i$ 分别代表第 $i$ 个拟合曲线的高度、中值和峰值锐度。用拟合后高斯函数的参数作为对目标光变曲线形态的量化参数。

## 3 程序实现

使用 Python 和 OpenCV 对算法进行实现，在 i7 4770k CPU 上测试，每幅图像的处理时间不超过 1 s，满足地基广角相机阵实时探测的需求。算法实现的关键信息如下:

(1) 异常图像过滤: Mini-GWAC 的单图像视场覆盖天区范围大，易受环境因素的影响。通过相邻图像相减后残差的方差和有效像素百分比这两个参数对图像进行过滤。

(2) 霍夫变换参数选择: 移动目标轨迹经过二值化之后，有可能形成多个线段。通过分析 Mini-GWAC 图像中流星的特征，选取概率霍夫变换的参数为 $\rho = 1$ 像素，$\theta = 1^\circ$，组成轨迹的子线段最小长度为 10 像素，子线段间最大间隔为 5 像素，轨迹的最短长度为 50 像素。

(3) 线段融合参数选择: 对图像中多条线段，如果中心距离小于 50，线段斜率差小于 5，且线段与图像中心的距离差小于 150，则这些线段合并为一条线段。

(4) 相邻帧线段集栈构造: 将连续多帧图像产生的线段集 $L(k)$ 缓存在一个堆栈中，用来过滤多帧出现的移动目标。相邻的两个线段集，对一个线段集中的所有线段都与另一个线段集中的线段进行比较，如果两个线段的斜率之差绝对值小于 5，则认为这两条线段属于同一个目标，即该目标是多帧出现的移动目标。

(5) 峰值数量 $N_{peak}$ 的计算: 首先对流量曲线进行中值滤波，滤波器的 kernelSize 选取目标长度的 10%。峰值计算过程 $Step_y = Flux_{max} / 10$，为过滤噪声带来的毛刺，限定相邻两个峰值的最大值之差应大于 $Step_y$。

(6) 光变曲线归一化: 将光变曲线的长度和高度(亮度)都归一化到 [0, 1]。对于所有光变曲线，记最大长度为 $L_{max}$，最大高度为 $H_{max}$，将所有光变曲线等比放大到长度为 $L_{max}$，对所有光变曲线的长度除以 $L_{max}$，完成长度归一化，对所有光变曲线的高度除以 $H_{max}$，完成高度的归一化。

(7) 拟合参数融合: 对于单峰目标的拟合，有些目标 $O_1$ 使用一个高斯函数能有效拟合，有些目标 $O_2$ 需要两个高斯函数才能有效拟合。对于 $O_1$，算法直接使用拟合的 3 个参数($A$，$x_0$ 和 $\alpha$) 描述其形态。对于 $O_2$，算法将两个高斯函数的参数进行融合，以更准确地描述流星曲线的形态: 对两个 $A$ 取最大值，对两个 $x_0$ 取均值，对两个 $\alpha$ 取最小值。

# 4 光变曲线形态特征分析

## 4.1 数据统计

对 Mini-GWAC 的 131 万幅图像进行分析处理，结果中各类目标的统计如表 2。

(1) 样本选择：这里选择单像素信噪比大于等于 5$\sigma$ 的移动目标，总数约 10.9 万个。

(2) 按是否多帧出现统计：包含 1.2 万个单帧出现的目标和 9.7 万个多帧出现的目标。

(3) 按光变曲线峰值个数统计：包含约 2.5 万个单峰目标、1.3 万个多峰目标和 7.1 万个无峰目标。

(4) 流星候选体：按照 Mini-GWAC 系统中流星的单帧出现特性对移动目标进行统计，流星候选体约占总数的 10.7%。如果通过光变曲线的单峰特性进一步过滤，则流星候选体的数据量可降低到 5.8%。

表 2 Mini-GWAC 不同类别的移动目标数量统计

Table 2 Statistic of different categories of moving objects in Mini-GWAC

| 统计量 | 数量/个 | 所占百分比/% |
|---|---|---|
| 移动目标总数 | 108 764 | 100 |
| 单帧出现数量 | 11 613 | 10.7 |
| 多帧出现数量 | 97 151 | 89.3 |
| 单峰目标数量 | 24 777 | 22.8 |
| 多峰目标数量 | 13 011 | 12.0 |
| 无峰目标数量 | 70 976 | 65.2 |
| 单帧出现单峰目标数量 | 6 275 | 5.8 |
| 单帧出现单峰峰值参数 $\alpha$ 大于 0.2 的目标数量 | 5 583 | 5.1 |
| 单帧出现单峰峰值参数 $\alpha$ 大于 0.2 且拟合度 $R^2$ 大于 0.7 的目标数量 | 4 453 | 4.1 |

为量化光变曲线的拟合精度，采用 R-Square ($R^2$) 进行评估。表 2 中的 6 275 个单帧出现单峰目标光变曲线高斯拟合结果 $R^2$ 的直方图如图 7。其中有 41.2%的目标 $R^2$ 大于 0.9；有 81.8%的目标 $R^2$ 大于 0.7。由于数据处理时存在噪声的干扰，所以选择较低的 $R^2$ 值(0.7)来过滤拟合结果。

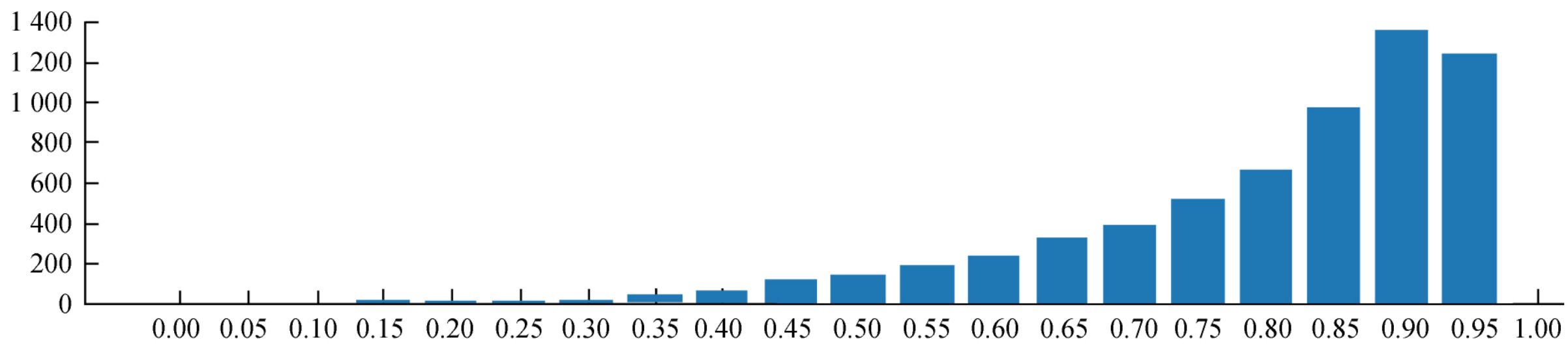

图 7 单帧出现单峰目标(6 275 个)光变曲线高斯拟合度 $R^2$ 直方图

Fig. 7 The histogram of $R^2$ of 6275 single-peak objects appeared in one frame

## 4.2 光变曲线形态分析

高斯拟合参数 $\alpha$ 代表移动目标光变曲线的波峰宽度：$\alpha$ 值越小，波峰越窄；$\alpha$ 值越大，波峰越宽。将基于 $\alpha$ 对单帧出现单峰目标的光变曲线进行统计分析。为了找到能在光变曲线形态上区分不同移动目标的参数，选择 3 个样本集，并对这 3 个样本集的波峰宽度值 $\alpha$ 进行直方图统计分析，这 3 个样本集分别是：

(1) 单帧出现单峰目标样本集，简称样本集 1：其中大部分为流星，且包含少量的多帧未匹配成功的空间碎片、卫星、飞机等长时间持续存在的移动目标，其波峰宽度值 $\alpha$ 分布如图 8。

(2) 傍晚或黎明期间的单帧出现单峰目标样本集，简称样本集 2：从(1)中选择晚上 9 点之前和凌晨 4 点之后出现的目标，其中的目标类别与(1)基本相同，其波峰宽度值 $\alpha$ 分布如图 9。

(3) 午夜期间的单帧出现单峰目标样本集，简称样本集 3：从(1)中选择晚上 10 点到凌晨 2 点之间出现的目标，该时间段的天空处于地球的阴影区，除了自发光的目标，很难观测到其他依靠反射太阳光才能看到的目标，飞机等自发光的目标受限于速度，多为多帧出现，单帧出现的概率非常低，因此该样本集中的目标主要为流星，其波峰宽度值 $\alpha$ 分布如图 10。

对这 3 个样本集 α 参数的直方图分布曲线分析如下：

(1) 分析图 8：样本集 1 的分布曲线有两个峰值。这两个峰值代表两类目标：一类是亮度随图像像素位置快速变化的目标，简称亮度快速变化目标，其 α 参数小于 0.2，占比 10%；一类是亮度随图像像素位置慢速变化的目标，简称亮度慢速变化目标，其 α 参数大于 0.2，占比 90%。

(2) 分析图 9：样本集 2 的分布曲线整体上与样本集 1 基本一致，但样本集 2 中快速变化目标的比例比样本集 1 高。

(3) 分析图 10：与样本集 1 和样本集 2 相比，样本集 3 的分布曲线在两端的比例偏少，在左边的快速变化部分也有一个很小的峰值，表明其中快速变化目标所占的比例非常少。样本集 3 中的移动目标主要为流星，因此流星中快速变化的个体非常少。

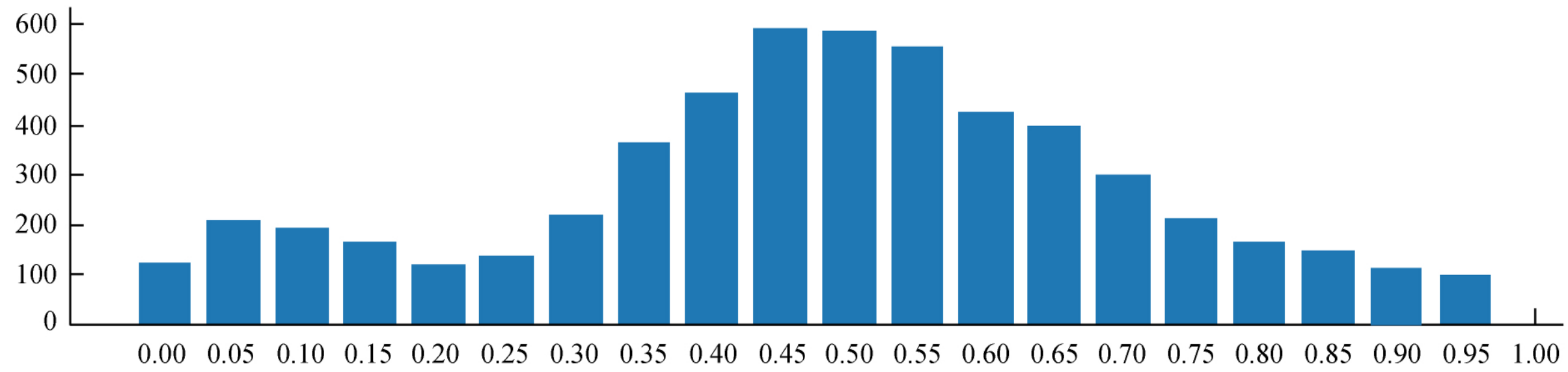

图 8 单帧出现单峰目标(6 275 个) α 参数的直方图

Fig. 8 The histogram of α parameter of 6275 single-peak objects appeared in one frame

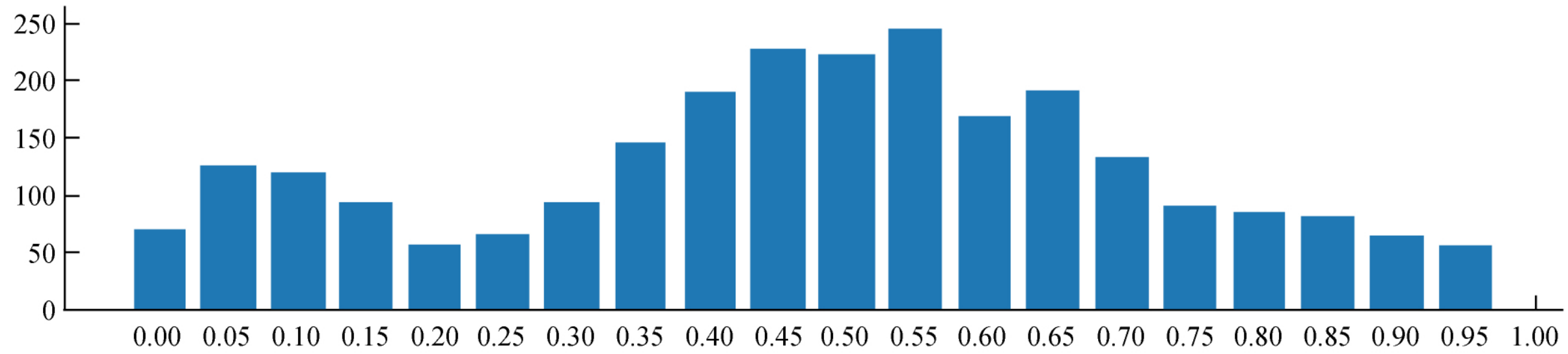

图 9 晚上 9 点之前和凌晨 4 点之后，单帧出现单峰目标(2 962 个) α 参数的直方图

Fig. 9 Before 9 p. m. and after 4 a. m, the histogram of α parameter of 2962 single-peak objects appeared in one frame

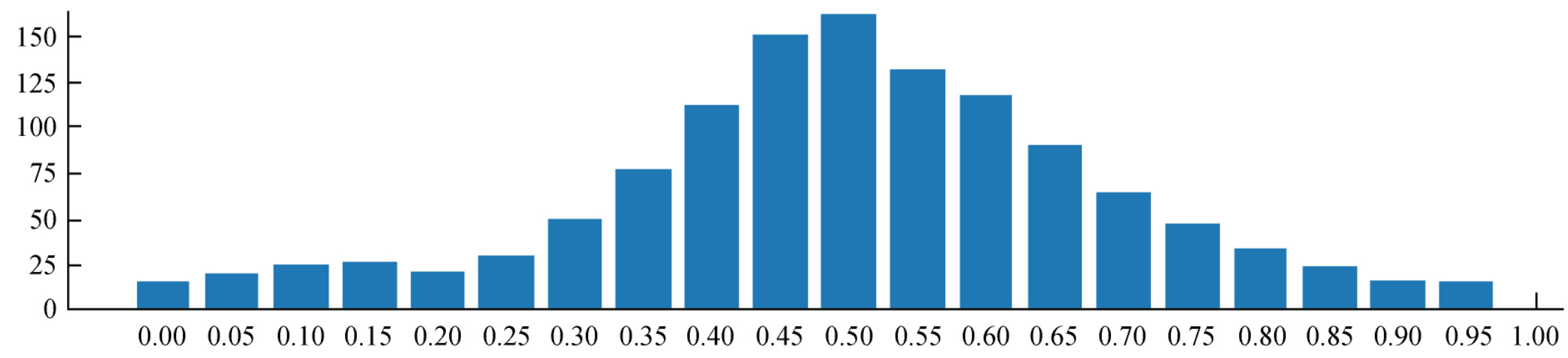

图 10 晚上 10 点到凌晨 2 点之间，单帧出现单峰目标(1 306 个) α 参数的直方图

Fig. 10 Between 10 p. m. and 2 a. m, the histogram of α parameter of 1306 single-peak objects appeared in one frame

综合分析图 8、图 9 和图 10 发现，大部分单峰值流星为亮度慢速变化的目标，依此条件对样本集 1 进行过滤，可以得到 5 583 个流星候选体，约占所有移动目标的 5.1%。同时采用 $R^2$ 大于 0.7 的条件过滤后得到 4 453(4.1%) 个目标，如表 2。人工对这 4 453 个目标进行检查，确认其中有 3 785 个目标符合流星的形态及亮度特征，占比 85%；有 104 个目标处于图像边缘，图像中仅包含目标的部分轨迹，这些部分轨迹也符合流星的形态及亮度特征，占比 2.3%；剩下 564 个为低信噪比目标、天气原因导致的假目标、与流星形态类似的天体目标等，占比 12.7%。因此算法的综合识别精度为 85%~87.3%。

## 5 总结与展望

地基广角相机阵的超大视场中每天探测到大量的流星，有效识别这些流星能为流星研究提供大量的有效数据。目前在地基广角相机阵这一类超大视场测光巡天系统中进行流星识别，遇到的最大挑战是没有充足的信息来有效区分流星与其他移动目标。为了尽可能准确地识别流星，设计和实现了一个面向地基广角相机阵这类系统的流星轨迹识别算法，同时对得到的流星候选体进行了光变曲线形态分析，结果总结如下:

(1) 识别算法对 Mini-GWAC 约两个月的历史图像进行处理，从中筛选出单像素信噪比大于 $5\sigma$ 的移动目标约 10.9 万个;

(2) 通过流星在 Mini-GWAC 图像中的单帧出现特性，可过滤 89.3%的移动目标，剩余 10.7%的候选目标;

(3) 大部分流星的光变曲线为单峰结构，该特性可进一步过滤候选目标，剩余 5.8%的流星候选体;

(4) 通过分析 3 个样本集光变曲线的 $\alpha$ 参数直方图，发现大部分流星为亮度慢速变化的目标。使用该特性和光变曲线拟合度 $R^2$ 进一步对候选目标进行筛选，最终得到 4.1%的流星候选体。

(5) 人工对 4.1%的流星候选体进行检查，确认其中有 85%~87.3%的目标符合流星的形态及亮度特征。

筛选流星使用了光变曲线单峰结构和亮度慢速变化这两个特性，这两个特征仅能包含大部分流星，而漏掉一些流星(如光变曲线有多个峰值的流星)。但是通过这两个特征的过滤，最终得到的 4.1%流星候选体具有更高的准确度。

通过分析单峰目标的实际图像发现，一些移动目标的形状和光变曲线形态与流星非常相似，如果这些目标出现在图像边缘，有很大概率仅单帧出现，本文算法很难过滤这类非流星目标。下一步将围绕 3 方面对算法进行改进: (1) 增加对移动目标的形态分析方法，如对移动目标的形状进行分析，找到多个可量化的形态描述参数，进一步提升流星的识别数量和识别精度; (2) 地基广角相机阵正在计划建设第 2 个站点，两个站点协同观测，可计算移动目标的高度信息，通过高度信息可过滤飞机和卫星等移动目标; (3) 正在研制快帧频 CMOS 相机，图像曝光时间可达到 0.1 s，将极大地增加移动目标的时间分辨率，可计算移动目标的速度信息。算法中加入高度和速度信息后，能更准确地区分流星与其他移动目标。

**参考文献:**

## An Algorithm of Selection of Meteor Candidates in GWAC System

Xu Yang[1,2], Wang Jing[1,2,3], Huang Maohai[1,2], Wei Jianyan[1,2]

(1. Key Laboratory of Space Astronomy and Technology, National Astronomical Observatories, Chinese Academy of Sciences, Beijing 100101, China; 2. University of Chinese Academy of Sciences, Beijing 100049, China, Email: yxu@nao.cas.cn; 3. Guangxi Key Laboratory for Relativistic Astrophysics, School of Physical Science and Technology, Guangxi University, Nanning 530004, China)

**Abstract:** With its large field of view, GWAC can record hundreds of meteors every day. These meteors are valuable treasures for some meteor research groups. It is therefore very important to accurately find all of these meteors. To address the challenge of precisely distinguishing meteors from other elongated objects in a GWAC-like sky survey system, we design and implement a meteor candidate recognition algorithm, including the recognizing and morphology analysis of the light curves of the meteor candidates. After processing the images of Mini-GWAC taken in two months, we detect 109000 elongated objects in which more than 90 percent of objects are not meteor. By analyzing $\alpha$ parameter of gaussian fitting curve of meteor target in midnight, we find that the wave crest of most single-peak meteor target changes slowly with the position of image pixel. Among the elongated objects, about 4.1% objects are identified as meteors with high confidence, after applying filters based upon an existence in a single frame, a single peak in the light curves, and a slow variation of the light curves. After manual examination, 85%-87.3% of the 4.1% candidates are consistent with the shape and brightness of the meteor.